\documentstyle[psfig]{aa}
\input epsf         
\begin{document}

\title{GRB\,050822:  Detailed analysis of an XRF observed by
  {\it\bfseries Swift}}

\author{O. Godet$^1$, K. L. Page$^1$, J. Osborne$^1$, B. Zhang$^2$, D. N.
  Burrows$^4$, P. T.  O'Brien$^1$, J. E.  Hill$^3$, J. Racusin$^4$, A. P.
  Beardmore$^1$, M. R.  Goad$^1$, A. Falcone$^4$, D. C. Morris$^4$, H.
  Ziaeepour$^5$}

\offprints{og19@star.le.ac.uk}

\institute{$^1$ X-ray and Observational Astronomy Group, Department of
Physics \& Astronomy, University of Leicester, LE1 7RH, UK\\
$^2$ Department of Physics, University of Nevada, Box 454002, Las
  Vegas, NV 89154-4002, USA\\
$^3$ NASA Goddard Space Flight Center, Greenbelt, MD 20771, USA\\
$^4$ Department of
  Astronomy \& Astrophysics, 525 Davey Lab, Pennsylvania State University,
  University Park, PA 16802, USA\\
$^5$ INAF-Osservatorio Astronomico di Brera, Via E. Bianchi 46, 23807, Merate (LC), Italy\\
$^5$ Mullard Space Science Laboratory, University College
  London, Holmbury St. Mary, Dorking Surrey, RH5 6NT, UK
}

\date{Received : / Accepted : }

\titlerunning{GRB\,050822} 
\authorrunning{Godet et al.}

\abstract{ 
  
  We report on the temporal and spectral characteristics of the early X-ray
  emission from the GRB 050822 as observed by {\it Swift}. This burst is
  likely to be an XRF showing major X-ray flares in its XRT light-curve.  The
  quality of the data allows a detailed spectral analysis of the early
  afterglow in the X-ray band. During the X-ray flares, a positive correlation
  between the count rate and the spectral hardness (i.e. the higher the count
  rate, the harder the spectrum) is clearly seen for the X-ray flares. This
  behaviour, similar to that seen for Gamma-ray pulses, indicates that the
  energy peak of the spectrum is in the XRT energy band and it moves towards
  lower energies with time.  We show evidence for the possible detection of
  the emergence of the forward-shock emission, produced at a radius larger
  than $4\times 10^{16}$ cm in the case of a CBM afterglow model (a formation
  region clearly different from that producing the prompt emission).  Finally,
  we show that the null detection of a jet break up to $T_0+4\times 10^6$\,s
  in the X-ray light curve of this XRF can be understood: i) if the jet seen
  on-axis is uniform with a large opening angle ($\theta > 20^\circ$); or ii)
  if the jet is a structured Gaussian-like jet with the line-of-sight outside
  the bright Gaussian core.

}

\maketitle

\section{Introduction}

X-ray flashes (XRFs) and X-ray rich Gamma-ray bursts (XRR GRBs) first detected
by {\it Ginga} and {\it BeppoSAX} (e.g. Heise et al.  2001) emit most of their
prompt energy in X-rays (see Lamb et al. 2004). It has been shown that XRFs
and GRBs share many observational properties, including: i) the temporal and
spectral properties of the prompt emission (e.g.  Heise et al.  2001, Kippen
et al.  2003, Sakamoto et al.  2005); ii) host galaxy properties (e.g.  Bloom
et al.  2003); iii) broadband afterglows as observed by {\it Swift} (Gehrels
et al.  2004 - XRFs 050215B, Levan et al. 2006a; 050315, Vaughan et al. 2006;
050406, Romano et al. 2006; 050416A, Mangano et al. 2006; 050714B, Levan et
al. 2006b; 060218, Campana et al. 2006b). The association of XRFs with
supernovae of type Ib/c (e.g. Tominga et al.  2004, Thomson et al.  2004,
Watson et al.  2004) suggests that XRFs and long GRBs share a similar
progenitor. Thus, it has been proposed that XRFs are simply an extension of
the long-GRB population with low values of the energy peak ($E_p$) of the
prompt spectra (e.g.  Sakamoto et al.  2005, Barraud et al.  2003).

A number of theoretical models have been proposed to explain XRFs.  Some are
based on intrinsic physical differences in the jet outflow (e.g. Mizuta et al.
2006) or in the jet geometries between XRFs and GRBs. Thus, the ``dirty
fireball'' invokes entrainment of baryonic material in the GRB jet,
resulting in a bulk Lorentz factor $\Gamma\ll 300$ (e.g.  Dermer et al. 1999,
Huang et al. 2002, Dermer \& Mitman 2004). Mochkovitch et al. (2004)
have alternatively proposed that GRB jets, in which the bulk Lorentz factor
$\Gamma> 300$ and the contrast between the bulk Lorentz factors of the
colliding relativistic shells is small, can also produce XRFs. It has also
been proposed that XRFs could simply have an intrinsically wider jet opening
angle in the case of a uniform jet model, since the energy peak of GRB
spectra is anti-correlated with the jet opening angle (Lamb et al. 2005; see
also Li et al. 2006).

On the other hand, other models simply invoke an effect of the viewing angle.
Indeed, M\'esz\'aros et al. (2002) have stressed that X-ray photons could be
produced by the view of the cocoon surrounding the GRB jet as it breaks out,
instead of the narrow jet (also see Zhang, Woosley \& Heger 2004).  Another
interesting model based on the unification scheme of AGN speculates that XRFs
could be the result of a highly collimated GRB jet viewed off the jet axis
(Yamazaki et al.  2003, Zhang et al. 2004).

The recent {\it Swift} broadband observations of XRFs have shown a variety of
temporal and spectral behaviour. In the case of the peculiar event 060218, it
was established that the explosion was quasi-isotropic (Soderberg et al.
2006).  Similarly, Mangano et al.  (2006) have shown that the jet opening
angle of GRB 050416A could be much larger ($\theta > 20^\circ$) than those
derived for GRBs ($5-10^\circ$, e.g. Frail et al. 2001).  On the other hand,
the light curve of GRB 050315 (Vaughan et al. 2006) shows evidence for a
possible jet break at $2.5\times 10^5$\,s, implying a jet opening angle of
$5^\circ$, consistent with the values derived for GRBs.  Finally, XRF 050406
has been shown to be a burst possibly seen well off the axis of a structured
jet (Schady et al. 2006).

These results suggest that the origin of XRFs is still not settled.  It is
therefore important to study the XRF-like events in detail to constrain their
true origins.

Here, we report the case of a burst detected by the {\it Swift} BAT (Burst
Alert Telescope; Barthelmy et al. 2005) on 22nd August 2005. The X-ray light
curve of this event exhibits a steep-to-flat-to-steep decay, and large X-ray
flares are superposed on the initial steep underlying decay. We show that the
spectrum of one of the X-ray flares could present a quasi-thermal component.
The paper is organised as follows: in Section 2, we present the
characteristics of the observations and the basic steps of the data reduction.
In Section 3, we present the temporal and spectral analysis of the
multi-wavelength observations. We establish that this burst is probably an XRR
GRB or an XRF, by using the BAT spectral results to compute the softness ratio
(e.g.  Lamb et al. 2004). In Section 4, we investigate the physical mechanisms
producing the spectral and temporal characteristics of the burst.

By convention, we note hereafter the flux in the X-ray band is modelled as
$F_\nu \propto \nu^{-\beta} (t-T_0)^{-\alpha}$, where $\beta$ is the energy
spectral index, $\alpha$ is the temporal index, and $T_0$ is the BAT trigger
time.  We use the symbol $\Gamma$ to refer to the bulk Lorentz factor. The BAT
spectral slope is noted as $\beta_{\mathrm{BAT}}$. All the time intervals are
hereafter referenced to the BAT trigger time.

\begin{table*}[h]
\begin{center}
\caption[]{Log of the XRT observations for GRB 050822 following the XRT mode sequence.}
\label{tab1a}
\begin{tabular}{ccccc}
\hline
\hline   
Sequence  & XRT Mode & Start time &  End time  & Start time since trigger\\
          &          &(yy-mm-dd hh:mm:ss) & (yy-mm-dd hh:mm:ss)  & (s) \\
\hline
00151486000 & IM & $2005-08-22 ~03:51:04$ & $2005-08-22 ~03:51:07$ & 95   \\
00151486000 & WT & $2005-08-22 ~03:51:20$ & $2005-08-22 ~03:55:22$ & 111     \\
00151486000 & PC & $2005-08-22 ~03:55:23$ & $2005-08-22 ~03:56:23$ & 354   \\
00151486000 & WT & $2005-08-22 ~03:56:26$ & $2005-08-22 ~03:59:45$ & 417   \\
00151486000 & PC & $2005-08-22 ~03:59:46$ & $2005-08-22 ~15:28:27$ & 617  \\ 
00151486001 & PC & $2005-08-22 ~16:23:49$ & $2005-08-23 ~00:54:34$ & 45260  \\ 
00151486002 & PC & $2005-08-23 ~01:01:53$ & $2005-08-23 ~17:08:59$ & 76344    \\
00151486003 & PC & $2005-08-24 ~01:05:26$ & $2005-08-24 ~23:40:58$  & 162957  \\
00151486004 & PC & $2005-08-25 ~11:41:57$ & $2005-08-25 ~23:35:57$  & 287548 \\
00151486005 & PC & $2005-08-27 ~20:35:26$ & $2005-08-28 ~23:34:13$  & 492357  \\
00151486006 & PC & $2005-08-30 ~03:17:39$ & $2005-08-30 ~22:42:58$  & 689290  \\
00151486007 & PC & $2005-08-31 ~07:49:39$ & $2005-08-31 ~22:44:58$  & 792010  \\
00151486008 & PC & $2005-09-01 ~11:22:21$ & $2005-09-01 ~21:18:57$  & 891172 \\
00151486009 & PC & $2005-09-02 ~00:26:17$ & $2005-09-03 ~22:44:13$  & 938208  \\
00151486010 & PC & $2005-09-04 ~19:48:16$ & $2005-09-06 ~04:03:57$  & 1180727 \\ 
00151486011 & PC & $2005-09-09 ~01:11:27$ & $2005-09-09 ~23:57:58$ & 1545718  \\ 
00151486012 & PC & $2005-09-11 ~04:44:06$ & $2005-09-12 ~22:42:58$ & 1731277 \\ 
00151486013 & PC & $2005-09-13 ~00:05:00$ & $2005-09-13 ~22:39:57$  & 1887331  \\ 
00151486014 & PC & $2005-09-15 ~00:05:59$ & $2005-09-15 ~22:53:31$  & 2060190\\ 
00151486015 & PC & $2005-09-17 ~00:13:10$ & $2005-09-18 ~06:52:57$  & 2233421 \\ 
00151486016 & PC & $2005-09-21 ~00:43:48$ & $2005-09-22 ~07:16:58$ & 2580859\\ 
00151486018 & PC & $2005-09-23~16:47:12$ &  $2005-09-26~23:59:58$  & 2822463 \\ 
00151486020 & PC & $2005-10-08 ~18:52:01$ & $2005-10-10~17:34:11$  & 4114952\\ 
00151486021 & PC & $2005-10-12 ~05:53:30$ & $2005-10-12~23:59:57$ & 4413841 \\ 
00151486022 & PC & $2005-10-13 ~01:32:24$ & $2005-10-13~03:21:58$ & 4484575  \\ 

\hline
\end{tabular}
\end{center}
\end{table*}

\section{Observation and data reduction}

\subsection{BAT observations}

The burst 050822 ({\it Swift}-BAT trigger 151486) was detected at 03:49:29
UT on 22nd August 2005 at (J2000) RA=
$03^{\mathrm{h}}24^{\mathrm{m}}19^{\mathrm{s}}$ and
Dec=$-46^{\mathrm{d}} 01' 22''$, with an uncertainty of 2 arc-minutes
(Blustin et al. 2005).

The BAT spectra and light curves were extracted using the BAT analysis
software (build 2.3) as described in the {\it Swift} BAT Ground
Analysis Software Manual (Krimm, Parsons \& Markwardt 2004).

\subsection{XRT observations}
\label{toto}

The X-Ray Telescope (XRT, Burrows et al. 2005) started to observe the burst
95\,s after the trigger, following the sequence of readout modes: Image mode
(IM) at the end of the slew, Windowed Timing (WT), and then Photon Counting
(PC) modes while pointed at the target (Hill et al. 2004, 2005). The Low Rate
Photo-Diode mode is no longer used, since the XRT CCD detector was damaged by
a micro-meteoroid on 27th May 2005 producing several bad columns (Abbey et al.
2005). The XRT observations are summarised in Table~\ref{tab1a}.  Note that
the X-ray light curve needed to be corrected for the loss of counts, because
the source is located on the CCD chip close to the bad columns. To do that, we
fitted the profile of the XRT point spread function (PSF, Moretti et al. 2005)
to estimate the fraction of lost counts in the IM data and the first orbit of
the WT and PC data (i.e. before $T_0+1000$\,s).  The correction factor applied
to the X-ray light curve before $T_0+1000$\,s is $f\sim 1.22$.

An uncatalogued X-ray source was identified at (J2000) RA =$03^{\mathrm{h}}
24^{\mathrm{m}} 27.26^{\mathrm{s}}$ and Dec =$-46^{\mathrm{d}} 02' 00.3''$
with an uncertainty of $1.4''$ at a 90\% confidence level. This refined
ground-calculated position was obtained after astrometry corrections.  To do
this, we remove the first 100\,s of each orbit where the star tracker attitude
was less stable.  For GRB 050822, this leaves only 216 ks of PC data.  The
data are further filtered to remove any remaining hot pixels that are not
filtered out by the normal pipeline processing, then exposure maps are made
based on the remaining data, and all images and exposure maps are summed.  We
obtain all of the optical objects within 15' from either SDSS if available or
USNO-B1 if not.  In the case of GRB 050822 we use USNO-B1.  To find
serendipitous X-ray sources for matching, we run {\scriptsize WAVDETECT} on
the combined XRT image, and then run {\scriptsize XRTCENTROID} to get the best
positions taking into account the instrument PSF and exposure maps.  We do not
do individual object to object matching, but rather we match all X-ray sources
to all optical sources and grab all matches with a separation of less than $20''$.
We look for clustering in those matches to find the overall mean frame shift.
We find the weighted mean frame shift measured from all the matches and remove
all outliers further away then 2-$\sigma$ from the mean.  We then iterate
finding the mean and removing outliers for a few more iterations also
requiring only one match per X-ray source on the third iteration. Finally, we
take this mean shift and apply it to the GRB position.  We calculate the
statistical position errors using the empirical fits as described in Moretti
et al. (2006), assuming that the astrometry correction removes the $3.5''$
systematic error normally applied to XRT positions to account for errors in
the star tracker attitude solution.  We add the statistical error to the error
from the frame shift due to the counting on each individual serendipitous
source. We note that our best XRT position is $0.5''$ away from the
astrometry corrected position (RA(J2000)=$03^{\mathrm{h}} 24^{\mathrm{m}}
27.22^{\mathrm{s}}$, Dec(J2000)=$-46^{\mathrm{d}} 02' 00.0''$ with an
uncertainty of $0.7''$ at 90\% confidence level) given for this burst by
Butler (2006).

The XRT data were processed by the {\it Swift} software version
2.5\,\footnote{See {\scriptsize
    http://heasarc.gsfc.nasa.gov/docs/swift/analysis/}}. This software
release includes new response files for the PC and WT modes which
significantly improves the spectral response at low energy (below 0.7 keV). It
is now possible to extend the fits down to 0.3 keV in both modes (Campana et
al.  2006a). The residuals below 0.6 keV are better than 10\% and the flux
accuracy between the PC and WT modes are better than 5\%.  A cleaned event
list was generated using the default pipeline, which removes the effects of
hot pixels and the bright Earth.  From the cleaned event list, the source and
background spectra were extracted using {\scriptsize XSELECT}.

Due to pile-up in the IM data, only the WT and PC data were useful for
spectral analysis.  The PC data from $\sim 354$\,s to $\sim 414$\,s and from
$\sim 617$\,s to $\sim 690$\,s with a count rate above 1 count s$^{-1}$ are
moderately piled-up. The innermost four-pixel radius was excluded, and the
source and background spectra were extracted using an annular region with an
outer radius of 20 pixels.  The same annular region was used to correct the
pile-up effect in the PC data for the X-ray light curve.  For the spectral and
temporal analysis, we used the grade 0-12 events for the PC mode and the grade
0-2 events for the WT mode, giving slightly higher effective area at higher
energies.  The ancillary response files for the PC and WT modes were created
using {\scriptsize XRTMKARF}.

\subsection{UVOT and other optical observations }

UVOT (UV-Optical Telescope; Roming et al. 2005), which began to observe 138\,s
after the trigger, detected no optical fading source down to a $3\sigma$
limiting magnitude of 19.5 in V-band for a 278\,s exposure and 19.4 in U-band
for a 188\,s exposure (Page et al.  2005).

ROTSE-III (Rykoff et al. 2005) started to observe 31.7\,s after the trigger
(i.e. during the Gamma-ray prompt emission phase), but no source was detected
down to an unfiltered magnitude of 16.6 in a 84\,s (at $T_0+31.7$ s) co-added
images exposure, and 17.5 in a 246\,s ($T_0+412.6$ s) co-added images
exposure.  

No redshift information is available for this event.

\section{Data analysis}

All the errors cited below are given at a 90\% confidence level for
one parameter of interest ({\emph{i.e.}} $\Delta\chi^2=2.706$).

\subsection{Spatial Analysis}

A faint X-ray source (SX) near the X-ray counterpart of GRB 050822 was
detected at (J2000) RA=$03^{\mathrm{h}} 24^{\mathrm{m}} 22.37^{\mathrm{s}}$
and Dec=$-46^{\mathrm{d}} 02' 10.55''$ with a 90\% error radius of $2.3''$
using the same method described in Section 2.2. This source is $4.9''$ away
from the X-ray counterpart of GRB\,050822.  A possible optical counterpart to
SX ($1.5''$ away from the SX position) is found in the catalogue USNO-B1.0
(RA(J2000) = $03^{\mathrm{h}} 24^{\mathrm{m}} 22.37^{\mathrm{s}}$,
Dec(J2000)=$-46^{\mathrm{d}} 02' 12.0''$; Monet et al.  2003).

\begin{figure}
\begin{center}
\hspace{0.8cm}\psfig{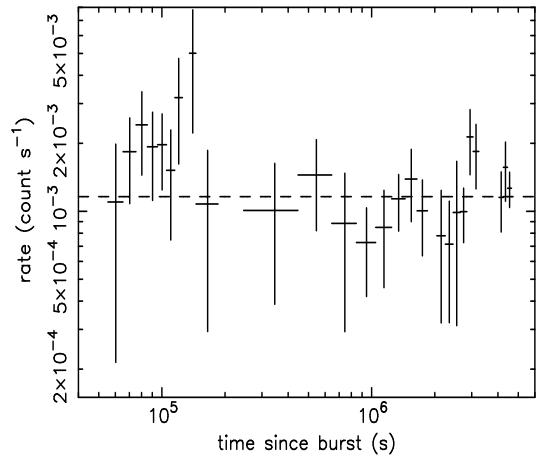}
\caption{0.3-10 keV X-ray light curve of the X-ray source close to the
  position of GRB 050822. The light curve is extracted from $\sim 5\times
  10^4$\,s to $\sim 5\times 10^6$\,s using a 15 pixel-radius extraction
  region.}
\label{figcr}
\end{center}
\end{figure}

Fig.~\ref{figcr} shows the X-ray light curve of this nearby source in the
0.2-10 keV energy band from $\sim 5\times 10^4$\,s to $\sim 5\times 10^6$\,s
using a 15 pixel-radius extraction region.  With a mean count rate of $\sim
10^{-3}$ counts s$^{-1}$, the source does not contaminate the light curve of
GRB\,050822 before $\sim T_0+3\times 10^5$\,s. While counts from GRB 050822
dominated the field, a 20 pixel radius was used to extract the light curve and
spectra. After $T_0+5\times 10^4$\,s, a 15 pixel extraction radius was used.

\subsection{Light curve}

\subsubsection{Gamma-ray band}
\label{Gamma}

\begin{figure}
\begin{center}
  \psfig{figure=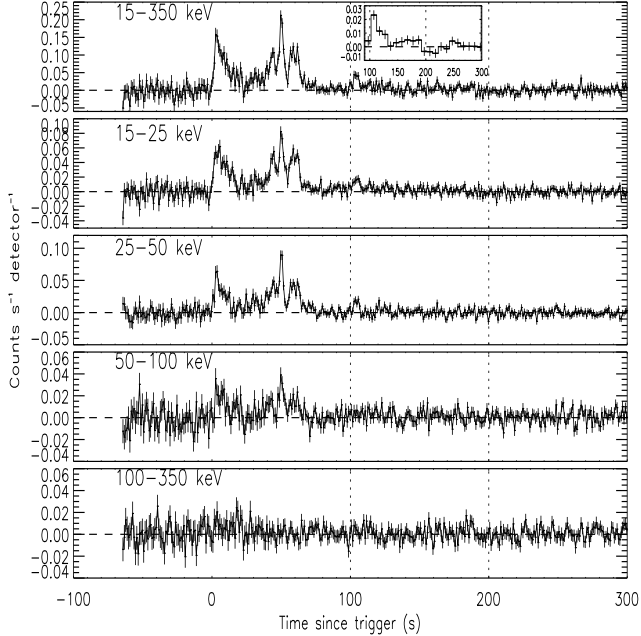,width=9.2cm,height=9.7cm}
\caption{Background-subtracted BAT light curve in units of count s$^{-1}$
  (fully illuminated detectors)$^{-1}$ for 4 different energy bands from the
  top to the bottom: 15-350 keV, 15-25 keV, 25-50 keV, 50-100 keV and 100-350
  keV.  The bin time is 1\,s. The dashed vertical lines delimit the temporal
  interval where the BAT and XRT data overlap.  In the small window using a
  bin time of 10\,s, we show the late small flare between 100\,s and 104\,s
  and a possible tail up to $T_0+200$\,s.}
\label{fig10}
\end{center}
\end{figure}

GRB 050822 exhibits a complex multi-peaked light curve (see Fig.~\ref{fig10}),
with peaks at $T_0+ \sim 0$\,s, $\sim 42$\,s, $\sim 48$\,s and $\sim
55-60$\,s. A small peak between $\sim 100$\,s and $\sim 104$\,s can be also
seen in the BAT light curve in the 15-25 keV and 25-50 keV energy bands. A
faint tail or flare from $\sim 104$\,s to $\sim 200$\,s can be also seen by
eye (see the small window in the top panel in Fig.~\ref{fig10}). Above 100
keV, only weak emission is seen in the BAT light curve.

$T_{50}$ and $T_{90}$ in the 15-350 keV band are $43.9\pm 0.2$\,s and
$104.7\pm 0.4$\,s, respectively.

\subsubsection{X-ray band}
\label{sect3}

\begin{figure*}
\begin{center}
\hspace{1.2cm}\psfig{figure=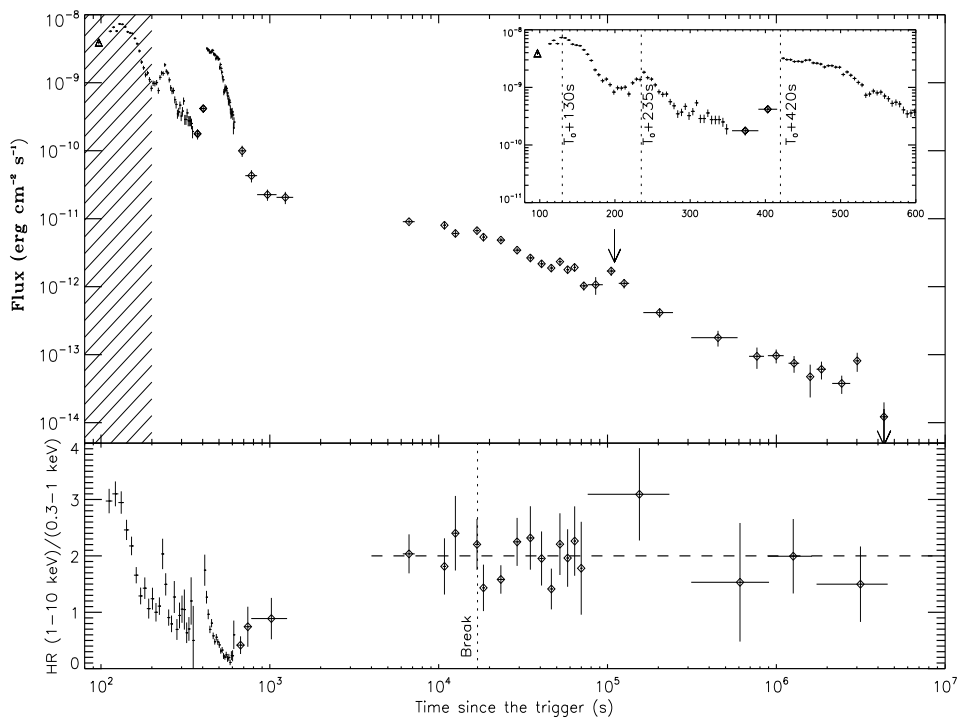,width=13cm,height=10cm}
\caption{Background subtracted XRT light curve of GRB 050822 in the 0.3-10 keV
  energy band in units of flux (top panel): IM data (triangle); WT 
  data (crosses); PC data (diamonds). The upper limits are given at 3
  $\sigma$. The times of the early X-ray flares are shown in the inset.
  The hatched area corresponds to the temporal BAT/XRT overlap.  The arrow
  around $1.1\times 10^5$\,s shows a possible late X-ray bump (see text for
  more details).  (Bottom panel) Hardness ratio of GRB 050822 of the 1-10 keV
  band over the 0.3-1 keV band as a function of time. The error bars are
  $1\sigma$ statistical errors. The diamonds correspond to the PC data and the
  crosses to the WT data. The vertical dotted line represents the break seen
  in the X-ray light curve (see Section 3.2.2). The dashed line is the mean
  value of the hardness ratio beyond $T_0+1000$\,s.  Note that for the HR
  plot, we did not include the piled-up PC data from $354$\,s to $414$\,s and
  from $617$\,s to $690$\,s.}
\label{fig2}
\end{center}
\end{figure*}

GRB\,050822 shows a complex XRT light curve in the 0.3-10 keV energy band (see
the top panel in Fig.~\ref{fig2}). The first 1000\,s of data display at
least three major X-ray flares peaking at $\sim 131$\,s (F$_1$), $\sim 236$\,s
(F$_2$) and $\sim 420$\,s (F$_3$) superposed on an underlying decay. 

Hereafter, we use the symbols F$_n$ to refer to these X-ray
flares. Note that the rise of the flare F$_3$ is relatively fast with a
timescale less than 30\,s. After $\sim 800-900$\,s, the light curve shows
a flat-to-steep decay similar to that seen in other {\it Swift} bursts. 

The bottom panel of Fig.~\ref{fig2} shows a hardness ratio, defined as the
ratio of the 1-10~keV band to the 0.3-1~keV band, as a function of time.  Some
spectral hardening and softening are clearly seen during the rising and
decaying parts of the X-ray flares, respectively, for the flares peaking
around $T_0+236$\,s (F$_2$) and 420\,s (F$_3$), and a clear spectral softening
is seen for the decay of the X-ray flare peaking at $T_0+131$\,s (F$_1$) (the
observation began during this flare).

We note that the X-ray and Gamma-ray emission during the temporal BAT/XRT
overlap are likely to be produced by the same mechanism (see Fig.~\ref{fig2}
and Section~\ref{specX}).  These clues suggest that the global decay seen
before 1000\,s could be the result of curvature effect emission (e.g.  Kumar
\& Panaitescu 2000 and Dermer 2004) associated with each X-ray flare. This is
confirmed by Liang et al.  (2006; see Figure 1 and also Table~1 in their
paper).

\medskip 

After 800-900\,s, the light curve can be described by a broken power-law with
an initial shallow slope $\alpha_1=0.45^{+0.12}_{-0.11}$ followed, after a
break at $1.7^{+0.5}_{-0.2}\times 10^4$\,s, by a steeper slope
($\alpha_2=1.05\pm 0.05$) with a possible late X-ray flare peaking at $\sim
1.1\times 10^5$\,s (see Fig.~\ref{fig2}).

\subsection{Spectroscopy}

\subsubsection{Gamma-ray band}

The BAT spectra are well fit by a single power-law. All the spectral
parameters and the fluence for different time intervals are summarised in
Table~\ref{tab2}.  The use of a Band function (Band et al. 1993) or a cutoff
power-law model did not significantly improve the fit.  The 15-150 keV fluence
GRB\,050822 is $2.3^{+0.2}_{-0.3}\times 10^{-6}$ erg cm$^{-2}$ over $T_{90}$,
which is moderate when compared to the average BAT fluence of $3.1\times
10^{-6}$ erg cm$^{-2}$ for GRBs from January 2005 to September 2006. The
spectral slopes of GRB\,050822 are relatively steep with respect to the
average BAT spectral slope of $\sim 0.8$ (e.g. O'Brien et al. 2006). This
suggests that $E_p$ may be below the BAT energy band, and that this burst may
be an XRF or an XRR GRB.

\begin{table}
\begin{center}
\caption[]{Summary of the Gamma-ray spectral fitting parameters.}
\label{tab2}
\begin{tabular}{ccccc}
\hline   
Time interval          & $\beta$ &  $\chi^2 (dof)$ & Fluence$^*$  \\
since $T_0$    &         &                 & ($\times 10^{-6}$ erg cm$^{-2}$) \\
\hline
0 - $T_{50}$      &  $1.36\pm 0.16$  & 47 (56) &  $1.4^{+0.2}_{-0.3}$\\
$T_{50}$ - $110$\,s&  $1.63\pm 0.25$  & 49.4 (56) &  $0.9\pm0.2$\\
0 - $110$ s       &  $1.46^{+0.15}_{-0.14}$  & 55.5 (56) &  $2.4^{+0.2}_{-0.3}$\\
\hline
90 - 125 s  & $1.57^{+0.73}_{-0.57}$ & 53.4 (56) & $0.2\pm0.1$ \\
\hline

\end{tabular}
    \begin{list}{}{}
      \item $^*$The fluences are given in the 15-150 keV energy range.
    \end{list}
\end{center}
\end{table}

Classification of a burst as an XRF or an XRR GRB depends on the softness
ratio of the 2-30 keV fluence over the 30-400 keV fluence (e.g. Lamb et al.
2004).  Bursts with $SR>0$ are classified as XRFs, bursts with $-0.5<SR<0$ are
classified as XRR GRBs, and those with $SR < -0.5$ are classified as normal
GRBs.  Because the BAT is not sensitive over this entire energy range, we
compute
$SR=\log\frac{S_X(2-30\,\mathrm{keV})}{S_{\gamma}(30-400\,\mathrm{keV})}$ by
integrating the best-fit spectra over these energy ranges.

If we assume that the energy peak $E_p$ is below 2 keV, we find $SR \sim
0.55^{+0.30}_{-0.49}$, corresponding to an XRF.  On the other hand, if we
assume $E_p=15$ keV and $\beta_{\mathrm{Band}}=0$ (which is the mean value of
the low energy spectral index of the Band function for GRBs and XRFs; Preece
et al.  2000, Kippen et al.  2003), we find $SR=0.12^{+0.08}_{-0.13}$, in the
XRF-XRR range.  A value of $\beta_{\mathrm{Band}}$ varying between
$-\frac{1}{3}$ and $\frac{1}{2}$ (i.e. the range of the low-energy spectral
index expected if the radiation is produced by the synchrotron mechanism; e.g.
Katz 1994) would still give a value of $SR > -0.5$.  So, the burst 050822 is
likely to be an XRF or an XRR GRB.

\subsubsection{X-ray band}
\label{specX}

The Galactic column density is $N_H^{Gal} = 2.34\,\times\,10^{20}$ cm$^{-2}$
in the direction of this burst (Dickey \& Lockman 1990). All the spectra were
binned to contain more than 20 cts bin$^{-1}$, and were fitted from 0.3 to 10
keV within {\scriptsize XSPEC} v11.3.1 (Arnaud 1996), except when the
statistics were too low, and in these cases, Cash statistics were used (Cash
1979). To model the absorption within {\scriptsize XSPEC}, we used the
photo-electric absorption model ({\scriptsize WABS}).

\smallskip

\begin{figure}
\begin{center}
\psfig{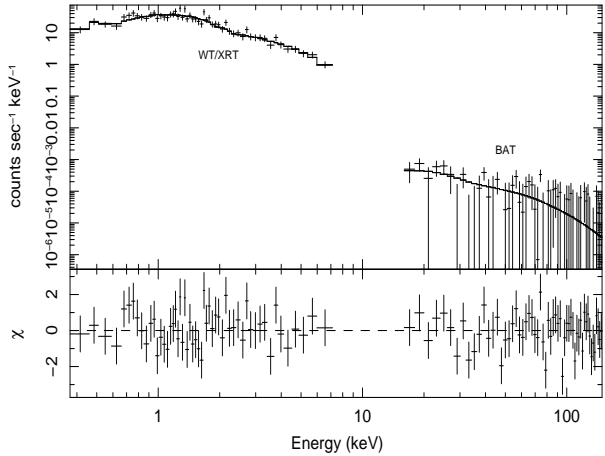}
\caption{Joint fit of the BAT/XRT spectra. The model is an absorbed power-law.}
\label{batxrt}
\end{center}
\end{figure}

{\it BAT/XRT analysis -} To investigate whether the early X-ray emission is
connected to the Gamma-ray emission, we fit the BAT and WT spectra from
$111$\,s to $125$\,s with an absorbed power-law. We did not use the BAT data
beyond $125$\,s because of poor statistics, and useful spectral data started
to be taken with the XRT only from $111$\,s (see Section 2.2).  A single
absorbed power-law gives a good fit with a slope of
$\beta=0.97^{+0.14}_{-0.13}$ and an excess absorption value of $\Delta n_H
(z=0)= 1.8\pm 0.4\times 10^{21}$ cm$^{-2}$ over the Galactic value
($\chi^2/\nu= 100/108$) using a constant factor ($f=0.95\pm 0.10$) to take
into account the difference in calibration between the XRT and the BAT (see
Fig.~\ref{batxrt}).  The use of an absorbed cutoff power-law or a broken
power-law did not significantly improve the fit.  This result is consistent
with the hypothesis that the X-ray and Gamma-ray emissions are produced by the
same mechanism during this time interval.

\medskip

{\it XRT analysis -} The fit of the WT data from 111\,s to 616\,s using an
absorbed power-law with absorption fixed to the Galactic value is poor with
$\chi^2/\nu = 1743/264$.  Leaving the absorption component free significantly
improves the fit with $\chi^2/\nu = 362/255$, with excess absorption of
$\Delta N_H (z=0)= 1.4\pm 0.1\times 10^{21}$ cm$^{-2}$ over the Galactic
value. Adding a black-body component further improves the fit by
($\Delta\chi^2=124$ for 2 DOF), and obtains an excess absorption column of
$\Delta N_H (z=0)= 9.5^{+2.0}_{-1.9}\times 10^{20}$ cm$^{-2}$.  A consistent
value of excess absorption ($\Delta N_H (z=0) = 1.2\pm 0.4\times 10^{21}$
cm$^{-2}$) is found when the PC data from $\sim 800$\,s to $\sim 4.1\times
10^4$\,s are fitted using an absorbed power-law.  Hereafter, we used two
{\scriptsize WABS} models in {\scriptsize XSPEC} with one fixed to the
Galactic value, and the other fixed at $1.2\times 10^{21}$ cm$^{-2}$, which is
the most reliable $\Delta N_H$-value because the late PC data are not affected
by spectral evolution (see the bottom panel in Fig.~\ref{fig2}).

We investigate the spectral evolution seen in the WT data by performing a
careful time-sliced spectral analysis of each flare. All the best-fit results
and the spectral models used in each case are summarised in Table~\ref{tab3}.
Note that the unabsorbed fluxes given in Table~\ref{tab3} are corrected for
the effect of the bad columns using a new tool {\scriptsize XRTEXPOMAP 0.2.1}
implemented in the version 2.4 of the XRT software.

\begin{table*}
\begin{center}
\caption[]{Summary of the X-ray spectral
  parameters for the best models. Note that we use two {\scriptsize WABS}
  models fixed to the Galactic value ($N_H^{Gal} = 2.34\,\times\,10^{20}$
  cm$^{-2}$) and to the value of excess absorption ($\Delta N_H(z=0)=1.2\times
  10^{21}$ cm$^{-2}$). The symbols F$_n$ correspond to the three X-ray flares
  peaking around $T_0+131$\,s (F$_1$), 236\,s (F$_2$) and 420\,s (F$_3$) as
  defined in Section 3.2.2.}
\label{tab3}
\begin{tabular}{llllcllc}
\hline   
XRT  & Time interval    & Model  & Flux$^\dagger$ &
$\chi^2 (\nu)$ &Model & Flux$^\dagger$& $\chi^2 (\nu)$  \\
 mode        & since $T_0$  &     &    &  &   &  &  \\
\hline
    &                &                      Power-law           &                      &    & & &           \\
 WT1(F$_1$) &  111 - 121\,s & $0.80\pm 0.10$         & $9.43^{+0.69}_{-0.65}$ & 21.2 (31) & & &   \\
 WT2(F$_1$) &  121 - 131\,s &  $0.79\pm 0.11$ & $8.46^{+0.60}_{-0.43}$ &  21.9 (28) & & &    \\
 WT3(F$_1$) &  131 - 145\,s &  $0.88\pm 0.08$ & $9.04^{+0.40}_{-0.49}$ & 42.4  (44) & & &   \\
 WT4(F$_1$) &  145 - 161\,s     & $1.08\pm 0.09$ & $6.86^{+0.35}_{-0.39}$ & 34.4 (39) & & &   \\
 WT5(F$_1$) &  161 - 181\,s                & $1.47\pm 0.13$ &
 $3.43^{+0.20}_{-0.18}$ &  36.8 (24) & & &   \\
 WT6(F$_1$) &  181 - 211\,s                   & $1.63^{+0.20}_{-0.19}$ &
 $1.61^{+0.11}_{-0.13}$ & 23.6 (16) & & &   \\
\hline 
    &                &                      Broken power-law             &
 &      &  & &         \\
    &                &                      $\beta_1=0$ fixed &
 &  &  & & \\
WT7(F$_2$) &  211 - 236\,s                 & $E_p<0.75$ keV & $1.47^{+0.14}_{-0.30}$ & 10.6 (12)  & & &  \\
    &                &               $\beta_2=1.72^{+0.23}_{-0.20}$      &
 &  & & & \\


 WT8(F$_2$) &  236 - 261\,s & $E_p=0.83^{+0.12}_{-0.15}$ keV     & $1.51^{+0.15}_{-0.16}$ & 5.3 (13) & & & \\
     &               & $\beta_2=1.88^{+0.33}_{-0.30}$     &                        &  & & & \\

WT9(F$_2$) &  261 - 301\,s  & $E_p<0.65$ keV                    & $0.69_{-0.62}^{+0.03}$  & 20.9 (17)$^\ddagger$ & & &\\
    &                & $\beta_2=2.02^{+0.48}_{-0.29}$             &                    &  & & & \\

\hline 
    &                &                      Power-law             &                      &      & & &         \\
 WT10(F$_2$) &  301 - 351\,s  & $1.99^{+0.23}_{-0.22}$ &
 $0.48^{+0.08}_{-0.07}$  & 109 (111)$^*$ & & &\\
\hline
    &                &                     Broken power-law              &    &    &  PL+BB$^a$ & &      \\
    &                &                      $\beta_1=1.06^{+0.16}_{-0.17}$ &                    &  & $\beta = 1.87^{+0.39}_{-0.62}$& & \\
    &                &                      $\beta_2=4.09^{+0.25}_{-0.21}$ &                    &  & & & \\
    &                &                      $E_p$ (keV)        & &      & kT (keV)& &         \\
 PC(F$_3$) &  355 - 415\,s  &  $ 2.63^{+1.08}_{-0.97}$  & $0.19\pm 0.04$ &  5.1 (9)$\ddagger$  &  & & \\
 
WT11(F$_3$) &  417 - 431\,s   & $1.74^{+0.29}_{-0.19}$  &  $2.11^{+0.18}_{-0.15}$ &
 181.1 (201) & $0.29\pm 0.04$ & $2.09^{+0.34}_{-0.23}$ & 159 (195)\\
 WT12(F$_3$) &  431 - 441\,s                  & $1.29^{+0.16}_{-0.11}$ &
 $2.38^{+0.19}_{-0.22}$ & - & $0.23\pm 0.03$& $2.48^{+0.42}_{-0.39}$ & -\\
 WT13(F$_3$) &  441 - 471\,s                  & $ 1.09^{+0.09}_{-0.09}$ & $2.27\pm
 0.13$ & -  & $0.19\pm 0.01$& $2.16^{+0.21}_{-0.12}$ &-\\
 WT14(F$_3$) &  471 - 501\,s                  & $ 0.91^{+0.05}_{-0.04}$  &
 $2.12^{+0.15}_{-0.13}$ & - & $0.17\pm 0.01$ &$2.00^{+0.24}_{-0.11}$ &- \\
 WT15(F$_3$) &  501 - 531\,s                  & $ 0.79^{+0.05}_{-0.04}$ &
 $1.68^{+0.15}_{-0.14}$  & - & $0.14\pm 0.01$ & $1.61^{+0.18}_{-0.16}$ &- \\
WT16(F$_3$) &  531 - 561\,s                  & $ 0.65\pm 0.05$  &
 $1.06^{+0.11}_{-0.15}$ & - & $0.12\pm 0.01$ & $1.06^{+0.15}_{-0.16}$& -\\
WT17(F$_3$) &  561 - 616\,s                  & $ 0.60\pm 0.06$  & $0.47^{+0.05}_{-0.08}$ & - & $0.11\pm0.01$& $0.50^{+0.07}_{-0.11}$& -\\
\hline
    &                &                      Power-law             &                      &        & & &       \\
 PC(F$_3$) &  625 - 789\,s                   & $3.12^{+0.62}_{-0.58}$  &
 $0.17^{+0.04}_{-0.03}$ & 15.8 (11)$^\ddagger$  & $0.08^{+0.04}_{-0.03}$$^b$& $0.16^{+0.08}_{-0.09}$ & 15 (10)$^\ddagger$\\

\hline
        &                                   & Power-law              &                      &       & & &        \\
 PC &  $\sim 800$ - $\sim 4.1\times 10^4$\,s             & $1.11\pm 0.09$ & $4.36^{+0.40}_{-0.40}\times 10^{-2}$ &  41.2 (47)      & & &  \\
 PC &  $\sim 800$ - $\sim 1.7\times 10^4$\,s      & $1.12\pm 0.15$ & $7.18^{+0.98}_{-1.03}\times 10^{-2}$  &  19.5 (20) & & & \\
 PC &  $\sim 1.7-4.1\times 10^4$\,s  &  $1.11\pm 0.11$ & $3.24^{+0.33}_{-0.37}\times 10^{-2}$ &  30.8 (26)& & &
 \\
 PC &  $\sim 8 - 12.5\times 10^4$\,s & $1.02^{+0.42}_{-0.37}$ & $9.46^{+4.69}_{-4.60}\times 10^{-3}$ &  6.1 (7)$^\ddagger$    & & &     \\
PC &  $\sim 0.4- 2\times 10^6$\,s   & $0.98^{+0.45}_{-0.42}$ & $6.87^{+3.87}_{-4.05}\times 10^{-4}$ &  62 (113)$^*$    & & &     \\
\hline

\end{tabular}
    \begin{list}{}{}
    \item $^\dagger$ $F$ is the unabsorbed flux given in the 0.3-10 keV energy
    band in units of ($\times 10^{-9}$ erg cm$^{-2}$).
    \item $^\ddagger$ Due to the low statistics, we used the Churazov
      weighting function within {\scriptsize XSPEC} when fitting the spectrum
      (see Churazov et al. 1996) and a binning of 10 counts per bin.
      \item $^*$ The Cash statistic value and the PHA bins in parenthesis.
      \item $^a$ PL+BB corresponds to a power-law plus black body model.
      \item $^b$ The spectral slope of the power-law component is fixed to
 $\beta=1.87$. 
    \end{list}
\end{center}
\end{table*}

The spectra for flare F$_1$ are well fit by an absorbed power-law with a
steepening of the spectral slopes with time. This is in agreement with the
hardness ratio (Fig.~\ref{fig2}). Since we have some evidence that the X-ray
and Gamma-ray emission is produced by the same radiation mechanism for a part
of flare F$_1$, we considered fitting the spectra with a Band function (Band
et al.  1993).  However, the narrow 0.3-10 keV XRT energy band does not allow
us to constrain the spectral parameters of the Band function with the
available statistics.  Instead, we approximated the Band function with an
absorbed broken power-law model\,\footnote{Note that within the {\scriptsize
    XSPEC} notations, the energy break ($E_b$) is one of the three parameters
  of the broken power-law model. We refer to $E_b$ in the text as the energy
  peak of the spectrum ($E_p$).}, where $\beta_1$ and $\beta_2$ are the low
energy and high energy spectral slopes, respectively.  Using that model does
not allow us to track the evolution of energy peak of the spectrum ($E_p$)
with time.  These points, along with the evidence that the X-ray and Gamma-ray
spectra from $111$\,s to $125$\,s are likely to be produced by the same
mechanism, suggest that $E_p$ has probably already passed through the XRT
energy band by $T_0+111$~s.

\begin{figure}
\begin{center}
  \psfig{figure=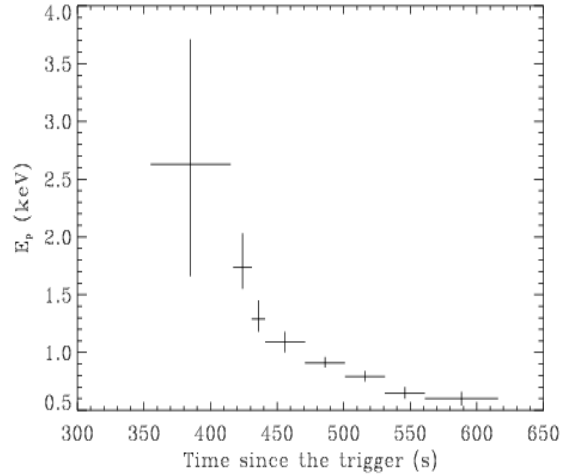,width=8cm,height=7cm,angle=0}
\caption{Evolution of the peak energy during flare F$_3$ as a function
  of time, for a broken power-law spectral model. }
\label{fig_spec}
\end{center}
\end{figure}

The WT spectra extracted for flare F$_2$ (from 211~s to 301~s post-burst) and
for flare F$_3$ (from $417$\,s to $616$\,s) are not well fit by an absorbed
power-law, since $\chi^2/\nu=42.9/29$ and $\chi^2/\nu=678/135$, respectively.
The use of a broken power-law for the spectra of flare F$_3$ allows us to
track the decrease of the peak energy with time (fitting both $\beta_1$ and
$\beta_2$, although they were tied to the same values for all the spectra -
see Table~\ref{tab3} and Fig.~\ref{fig_spec}).  Although $E_p$ was well
determined during the bulk of this flare, it was not constrained at all for
the PC spectrum from $625$\,s to $789$\,s at the end of the flare.  When this
spectrum is fit by a single absorbed power-law, the spectral slope is still
inconsistent with the $\beta_2$-value derived from the earlier WT spectra.
The spectral hardening seen after $\sim T_0+700$\,s (see Fig.~\ref{fig2})
indicates that an extra X-ray emission component is probably present at this
time and may account for the inconsistency in the $\beta$ value (see Section
4.3).  For flare F$_2$, even assuming that $\beta_1=0$ (i.e. the mean Band
function low energy spectral slope for BATSE bursts), we obtained only upper
limits for $E_p$ for spectra WT7 and WT9, the spectrum WT10 being best fit by
a single absorbed power-law with $\beta \sim 2$. This $\beta$ value is
consistent with the $\beta_2$-values of the broken power-law model (see
Table~\ref{tab3}).  Although we were able to measure $E_p$ only for spectrum
WT8, it is not completely clear whether $E_p$ varies during this flare. The
results suggest that, like for flare F$_1$, $E_p$ was below the XRT energy
band for most of flare F$_2$.

We showed that the average spectrum of flare F$_3$ (from 417\,s to 616\,s) can
alternatively be fitted using a black-body (BB) $+$ power-law (PL) model,
which was previously used in the case of GRB 060218 (Campana et al.  2006b),
with $kT=0.185^{+0.007}_{-0.006}$ keV, $\beta=2.08^{+0.15}_{-0.17}$, and
$\chi^2/\nu=147/133$ . The BB flux corresponds to $62.5\pm 3.9\%$ of the total
observed 0.3-10 keV flux. The use of that model allows us to track the
decrease of the BB temperature with time. There are also some hints that the
power-law component steepens with time; however, the values of the spectral
slope for each spectrum are consistent within the error bars, probably due to
the poor statistics of the spectra at later times.  The value of
$\beta=1.87^{+0.39}_{-0.62}$ in Table~\ref{tab3} was obtained by tying the
spectral slope to the same value for each spectrum.  The PC spectrum from
$\sim 355$\,s to $\sim 415$\,s is well fitted with a single power-law with
$\beta=1.62^{+0.37}_{-0.35}$ ($\chi^2/\nu=5/9$); adding a BB component for
this spectrum does not improve the fit significantly ($\Delta\chi^2 = 1.4$ for
2 dof).  For the PC data from $\sim 625$\,s to $\sim 789$\,s, if we fix the
spectral slope of the power-law to $\beta=1.87$, then we can constrain the
temperature of the black body (see Table~\ref{tab3}).  The use of a
{\scriptsize BBODYRAD} model allows us to constrain the X-ray emitting radius
of the BB component, which increases from $R_X^i\sim 2.6^{+0.9}_{-0.6} \times
10^{13} \times \left(\frac{D_L}{20\,\mathrm{Gpc}}\right)$~cm at the time
interval 417-431\,s to $R_X^f\sim 9.8^{+1.8}_{-1.9} \times 10^{13} \times
\left(\frac{D_L}{20\,\mathrm{Gpc}}\right)$~cm at the time interval 471-501\,s,
where $D_L$ is the luminosity distance.  We normalise to $D_L=20$~Gpc, the
luminosity distance for the average redshift ($z\sim2.5$) for {\it Swift}
GRBs.  At later time, the emission radius is no longer well constrained. The
results are summarised in Table~\ref{tab444}.

\begin{table}
\begin{center}
\caption[]{Summary of the emitting radius of the black-body component for the
  WT data fit from 417\,s to 501\,s.}
\label{tab444}
\begin{tabular}{cc}
\hline   
Time interval  & $R_X$  \\
since $T_0$    & ($\times 10^{13}~D_{20\,\mathrm{Gpc}}$ cm) \\
\hline \vspace{-3mm} \\
 417-431\,s &  $2.6^{+0.9}_{-0.6}$ \vspace{1mm}\\
 431-441\,s &  $4.4^{+1.2}_{-1.5}$ \vspace{1mm}\\
 441-471\,s  & $6.9^{+0.8}_{-0.7}$ \vspace{1mm}\\
 471-501\,s  & $9.8^{+1.8}_{-1.9}$ \vspace{1mm}\\
\hline
\end{tabular}
    \begin{list}{}{}
    \item $^*$ $D_{20\,\mathrm{Gpc}}=\frac{D_L}{20\,\mathrm{Gpc}}$, where
      $D_L$ is the luminosity distance of the source and 20~Gpc is the
      approximate luminosity distance of the mean {\it Swift} GRB
      redshift, using WMAP cosmology.
    \end{list}
\end{center}
\end{table}

\bigskip

The time sliced-analysis of the PC data beyond 800\,s reveals that no
significant spectral variation is seen around the break in the light curve at
$\sim 1.7\times 10^4$\,s or the late X-ray bump around $1.1\times 10^5$\,s
(see Table~\ref{tab3}).

\section{Discussion}

We established in Section 3 that the burst 050822 is an XRF or an XRR GRB. Its
X-ray light curve shows a steep-to-flat-to-steep decay. At least three X-ray
flares peaking around $T_0+131$\,s (F$_1$), 236\,s (F$_2$) and 420\,s (F$_3$)
are superposed on the initial steep decay.  A strong spectral evolution is
observed during the flares. Flares F$_2$ and F$_3$ are best fit by broken
power law spectral models, and we showed that the spectral softening during
the decaying part of the flares is probably due to the shift of the energy
peak of the spectrum ($E_p$) to lower energies.  Interestingly, we found that
the data for the X-ray flare F$_3$ are also well fitted by a black-body plus
power-law model as in the case of GRB 060218, the black-body component cooling
down and expanding with time.

We discuss the possible origin of the X-ray flares in the framework of the
internal shock model, which is often invoked to interpret the such flares
(e.g.  Zhang et al. 2006, King et al. 2005).

The X-ray light curve also shows a long smooth decay from $\sim 1.7\times
10^4$\,s to $\sim 4\times 10^6$\,s without any evidence for a jet break. We
investigate whether it is consistent with the prediction of the current
afterglow models. We also discuss the possible origins of the late X-ray bump
around $T_0+1.1\times 10^5$\,s and its implications.

\subsection{The origin of the X-ray flares}

The two early X-ray flares peaking around $T_0+236$\,s (F$_2$) and 420\,s
(F$_3$) clearly show a positive correlation between brightness and spectral
hardness i.e. the higher the count rate, the harder the spectrum (see the
bottom panel in Fig. 4), as found for instance by Ford et al. (1995) in GRB
pulses. We also note that the temporal profiles of the X-ray flares are well
fitted by common FRED pulse shape (see Liang et al. 2006).  The spectral
softening seen for the X-ray flares can be explained by a shift of the energy
peak to lower energy through the XRT energy band. This is clearly seen for the
X-ray flare F$_3$, for which the data are well fit by an absorbed broken
power-law (see Fig.~\ref{fig_spec} and Table~\ref{tab3}).  Indirect evidence
for the shift of the peak energy to lower energies is also presented for the
two other X-ray flares in Table~\ref{tab3}.

\medskip
{\it Internal shocks -} The presence of the energy peak in the XRT energy band
is consistent with the internal shock model, as shown by Zhang \& M\'esz\'aros
(2002). In this model, the peak energy $E_p$ of the synchrotron emission
satisfies:
$$E_p\propto L^{1/2}~\Gamma^{-2}~\delta t^{-1}$$
where $L$ and $\delta t$ are
the luminosity and the variability timescale, respectively.  A smaller
luminosity and/or a higher value of $\Gamma$ and $\delta t$ produces X-ray
flares rather than Gamma-ray peaks. Here, the $\delta t$-values of the flares
are larger than those of the Gamma-ray peaks. A longer duration of the X-ray
flares is indeed expected at later times due to longer accretion episodes
around the central new-born compact object (e.g. Perna et al.  2005 and Proga
\& Zhang 2006).  It is not completely clear if the late $\Gamma$ are higher or
not.  We could speculate that the late ejected shells interacting with a
cleaner environment along the jet axis have higher $\Gamma$. It is
nevertheless more likely that the main factor to produce a lower $E_p$ is a
smaller luminosity at later time.  The shift of $E_p$ in X-ray flares through
the XRT energy band has also been reported in other {\it Swift} bursts (e.g.
GRB 051117A, Goad et al.  2006).  These authors also concluded that the X-ray
flares are produced by internal shocks.

The low energy spectral slope ($\beta_1= 1.06^{+0.16}_{-0.17}$) for the X-ray
flare F$_3$ is steep compared to the mean values of the $\beta_1$ distribution
derived from a sample of averaged time GRBs and XRFs ($\beta_1\sim 0$; see
Preece et al. 2000 and Kippen et al. 2003).  The steep $\beta_1$-value could
suggest that the X-ray emission is not produced solely by synchrotron
radiation, since the low energy spectral slope from shock accelerated
electrons is expected to be between $-\frac{1}{3}$ and $\frac{1}{2}$ (e.g.
Katz 1994, Cohen et al.  1997, Lloyd \& Petrosian 2000).

\bigskip

\begin{figure}
\begin{center}
\psfig{figure=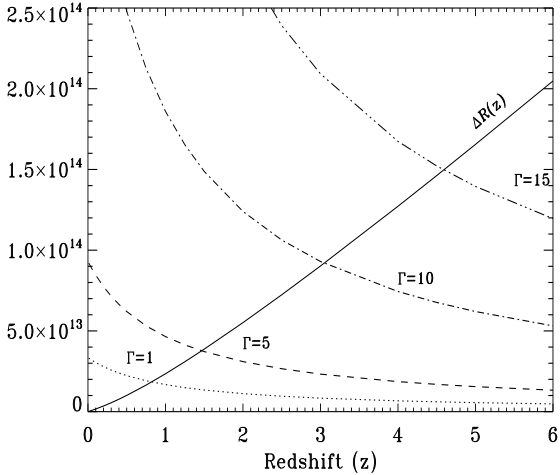,width=8cm,height=7cm}
\caption{Evolution of the two parts of Eq.~1 as a function of the redshift (z)
  assuming that the Lorentz factor $\Gamma$ has reached its coasting value.
  The intersections between the dotted and thick lines indicate the possible
  $z$-solutions for Eq.~1. The two parts of Eq.~1 are expressed in units of
  cm.}
\label{fig_zz}
\end{center}
\end{figure}

{\it Photospheric emission -} In Section 3.3.2, we showed that the spectra of
the X-ray flare F$_3$ are alternatively well fit by a black-body plus
power-law model.  According to the internal shock model, a quasi thermal
spectrum is expected to be produced by pair photospheric emission from an
optically-thick shocked shell of matter becoming optically thin at a radius
$R_{\tau}$.  However, Comptonisation of the photospheric emission during the
emergence of the spectrum (Goodman 1986, Ryde et al. 2006, Thompson et al.
2006) or a strong magnetic component could lead to a non-thermal tail in the
spectrum (Thompson 1994; M\'esz\'aros \& Rees 2000; M\'esz\'aros et al. 2002;
Rees \& M\'esz\'aros 2005). It is difficult to know whether or not the
spectrum would peak in the XRT band, since it depends on the pair optical
depth and the pair temperature (e.g.  Zhang \& M\'esz\'aros 2002). We note
that Pe'er et al.  (2006) concluded that energy peaks below a few keV are not
expected in that picture. However, the parameters used in that paper were for
the prompt emission. It is possible that for plausible X-ray flare
parameters, the photospheric thermal component may be as low as keV or less
(although more detailed modeling is needed, which is outside the scope of our
paper). 

Assuming that the shell of matter moves relativistically, the variation of the
emission radius in the ``thin shell'' case is given by:
$$\Delta R(z) \equiv R_X^f(z) - R_X^i(z) =
\frac{2\,c}{(1+z)}[t_f~\Gamma^2(t_f) - t_i~\Gamma^2(t_i)]~~(1)$$
where $z$ and
c are the redshift and the velocity of light, respectively. Here, we define
$t_{i,f}$ as the mean times of the time intervals 417-431\,s and 471-501\,s
respectively. Fig.~\ref{fig_zz} shows the two parts of Eq.~1 as a function of
the redshift assuming that the Lorentz factor $\Gamma$ has reached its
coasting value. From the two plots, it seems more likely that the shell is
mildly relativistic ($\Gamma < 10-15$).  Otherwise, the solution of the
above equation would require an unreasonable high redshift ($\Gamma=100$ would
require a redshift much larger than 6).

\subsection{The X-ray light curve before $T_0+800$\,s: the tail of the prompt emission}
\label{prompt}

It is likely that the X-ray light curve before $T_0+800$\,s is associated with
the tail of the prompt emission: i) the X-ray flares are likely to be produced
by internal processes; ii) curvature effect emission associated with
the X-ray flares can account for the underlying decay seen in the X-ray
light curve before $T_0+800$\,s (see Section 3.1.2 and Liang et al. 2006);
iii) the X-ray and Gamma-ray spectra from $T_0+111$\,s to $T_0+125$\,s are
likely to be produced by the same physical mechanism (see Section 3.2.2).

\subsection{Constraints on the evolution of the afterglow}
\label{afterglow}

{\it The afterglow emergence -} It is worth noting that a spectral hardening
with time is clearly seen after $\sim T_0+700$\,s in the WT and PC data (see
the bottom panel in Fig.~\ref{fig2}).  Evidence that the X-ray continuum
emission is sometimes harder during the shallow decay of the XRT light curves
than during the initial steep decay has been found in several {\it Swift}
bursts (e.g.  O'Brien et al. 2006).  The X-ray emission producing the initial
steep decay and that producing the flat-to-steep decay were then interpreted
as arising from different mechanisms (i.e.  processes associated with the
prompt emission, as discussed in Section~\ref{prompt}, and external forward
shock, respectively).  We argue here that the spectral hardening seen after
$\sim T_0+700$\,s could be interpreted as the emergence of the forward-shock
emission.  In the case of the ``thin shell'' CBM (Circum Burst medium), we
could calculate a lower limit on the Lorentz factor from Eq. 10 in Zhang et
al. (2006):
$$\Gamma \ge 100
\left(\frac{t_{\mathrm{dec}}}{180\,\mathrm{s}}\right)^{-3/8}
E^{1/8}_{iso,52}\left(\frac{\eta}{0.2}\right)^{-1/8}n^{1/8}~\left(\frac{1+z}{2}\right)^{3/8}(2)
$$
where $t_{\mathrm{dec}}$ is the deceleration time (here
$t_{\mathrm{dec}}\sim 700$\,s).
$E_{iso,52}=\frac{E_{iso}}{10^{52}\,\mathrm{erg}}$, $\eta$ and $n$ are the
isotropic energy of the burst, the efficiency for the conversion of kinetic
energy into gamma-rays and the CBM density, respectively. From the work of
Sakamoto et al. (2005, 2006), it appears that XRFs follow the Amati (Amati et
al.  2002) relation (see also Lamb et al. 2005; Amati et al.  2007).  Assuming
that the burst follows this relation and $E_p< 15$ keV (see Section 3.2.1),
the isotropic energy should be less than $E_{iso} < 3\times 10^{50}~(1+z)^2$
erg.  Since $\Gamma$ in Eq. 2 depends weakly on $\eta$ and $n$, we obtain a
value of $\Gamma\ge 30\times (1+z)^{5/8}$. This would give a minimum
deceleration radius of $R_{\mathrm{dec}}\sim 2~c~t_{\mathrm{dec}}
\frac{\Gamma^2}{(1+z)} \ge 4\times 10^{16}~(1+z)^{9/8}$ cm. The inferred
radius is much larger than the radius usually thought of for the production of
the internal shocks ($10^{13}-10^{14}$ cm). The site of the emission after
$T_0+700$\,s is then likely to be different to that producing the emission
before $T_0+700$\,s according to the CBM afterglow model.

\medskip

{\it Standard afterglow model -}   If the blast-wave evolution has
already entered the slow cooling regime when deceleration started (i.e. $\nu >
\mathrm{max}(\nu_m,\nu_c)$ where $\nu_m$ and $\nu_c$ are the synchrotron and
cooling frequency respectively), then the temporal decay index ($\alpha$) and
the spectral slope ($\beta$) after the break at $\sim 1.7\times 10^4$\,s, are
predicted to be $\alpha=(3p-2)/4$ (a) and $\beta=p/2$ (b) according to the CBM
model (e.g.  Sari et al.  1998) and the wind model (e.g.  Chevalier \& Li
2000), where $p$ is the power law index of the electron distribution.  We find
$p=2.22\pm 0.18$ using (b) and $p=2.07\pm 0.07$ using (a). These values are
consistent with the commonly used values of $p=2.0-2.4$ (e.g. Kirk et al.
2000, Achterberg et al. 2001).

The shallow decay from $\sim T_0+800$\,s to $\sim 1.7\times 10^4$\,s can then
be interpreted as a phase of energy injection in the blast-wave, possibly due
to a longer activity phase of the central engine (such as the kinematic
luminosity $L\propto t^{-q}$) or a wide distribution of ejecta Lorentz factors
(Rees \& M\'esz\'aros 1998; Zhang \& M\'esz\'aros 2001; Nousek et al. 2006;
Zhang et al. 2006). In the case where $\nu > \mathrm{max}(\nu_m,\nu_c)$, we
find $q=0.32\pm 0.15$, which is consistent with previously determined
$q$-values (e.g. Zhang et al.  2006).

\begin{figure}
\begin{center}
  \hspace{-0.8cm}\psfig{figure=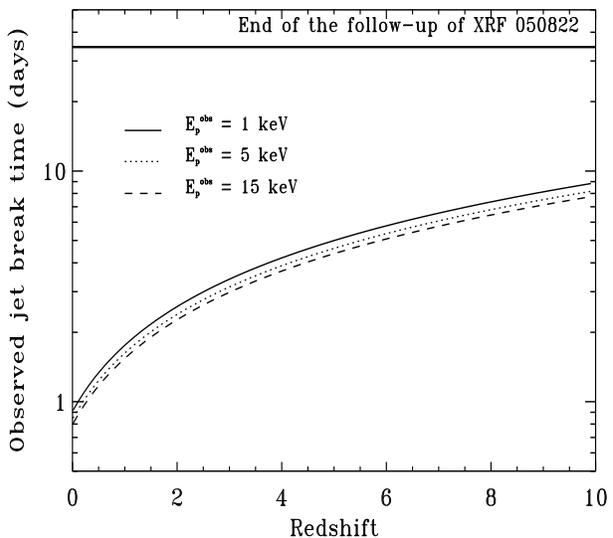,width=9cm,height=8.0cm}
\caption{Evolution of the jet break time as a function of the redshift for
  three values of the observed peak energy $E^{obs}_p$. The curves are deduced
  from the combination of the Amati et al. (2002) relation and Eq. 5 in Liang
  \& Zhang (2005). The thick, solid line corresponds to the end of the XRT
  follow-up.}
\label{fig_jet}
\end{center}
\end{figure}

\subsubsection{Any evidence for a jet break ?}

The X-ray light curve of GRB 050822 after $T_0+1.7\times 10^4$\,s shows a
monotonic, relatively smooth (except the late bump around $1.1\times 10^5$\,s)
and long decay up to $T_0+4\times 10^6$\,s.  No indication of any jet break is
seen.

Fig.~\ref{fig_jet} shows the expected observed jet break time ($t_{jet}$) for
different values of the observed energy peak ($E_p^{obs}$) as a function of
the redshift $z$, using the relations from Amati et al. (2002; A02) and Liang
\& Zhang (2005; LZ05). Note that the LZ05 relation was originally established
for optical breaks.  However, if the jet models are correct, then the jet
break time in the X-ray band should be the same.  From the figure, it appears
that whatever the values of $z$ and $E_p$, a jet break is expected in the
light curve within the first 10 days after the burst.  No such break is seen.
A similar result was found in GRB 050416A (an XRF; Sakamoto et al. 2006), for
which the A02 and LZ05 relations were inconsistent with the lack of a jet
break up to $T_0+34.5$ days.

We discuss in the next Section whether the apparent absence of a jet break in
the light curve can be understood in the framework of the current jet models.

\subsubsection{The jet models}

The model of the off-axis uniform jet with the line-of-sight outside the jet
edge (e.g. Yamazaki et al. 2003) can be ruled out.  Indeed, this model
predicts an initial fast rise when the emitting surface enters the
line-of-sight of the observer followed by a rapid decay with $\alpha\sim p$
(e.g.  Granot et al.  2002, 2005). This model is inconsistent with our data.

The model of the two-component jet with the line of sight on or close to the
less energetic wider beam is also not favoured. In such a model, it is
expected that an afterglow rebrightening is seen when the fireball is
decelerated so that the more energetic narrow component enters the field of
view. The lack of any significant rebrightening feature suggests that the
distinct two jet components as required by the model are not needed.

GRB\,050416A also has a very long power-law decay in its X-ray light curve,
with no indication of a jet break.  Mangano et al.  (2006) have modelled the
X-ray light curve of GRB\,050416A using two jet models: (1) an on-axis uniform
jet with a very wide opening angle (e.g. Lamb et al.  2005); (2) a structured
Gaussian-like jet with the line of sight outside the bright Gaussian core
(Zhang et al. 2004). We can infer from their Figure 5 that either of these jet
models could work in the case of GRB 050822.  In the case of the on-axis
uniform jet model, the lack of a jet break in GRB\,050822 requires a large jet
half-opening angle (up to $\theta > 20^\circ$).

\subsubsection{Origin of the late X-ray bump around $T_0+1.1\times 10^5$\,s}

We next consider whether the X-ray bump around $t_{\mathrm{bump}} =
T_0+1.1\times 10^5$\,s could be produced by external shocks.  Indeed, it has
been proposed that abrupt density fluctuations in the circumburst medium can
produce a significant re-brightening in the GRB afterglows via external shocks
(e.g.  Lazzati et al.  2002).  However, if the blast-wave is still in the
relativistic regime, the flux at $\nu > \nu_c$ should not (or only very
weakly) be affected by circumburst density fluctuations (e.g. Nakar et al.
2003). Recent work has shown that if the blast-wave is still in the
relativistic regime after the interaction and $\nu>\nu_c$, then the decay
slope $\alpha$ is expected to vary, but no re-brightening is expected to be
seen in the X-ray light curves (Nakar \& Granot 2006). So the bump around
$1.1\times 10^5$\,s is unlikely to be produced by the result of the
interaction of the blast-wave with some clouds of matter or density jumps.

\smallskip

>From the quality of the data around $t_{\mathrm{bump}}$, we could not
completely rule out that the X-ray bump may be produced by an inhomogeneity in
the blast-wave or by energy injection when we compare the rising ($ 0.2<
\delta t_r/t_{\mathrm{bump}} < 1$) and decaying timescales ($\delta
t_d/t_{\mathrm{bump}} \sim 1$) of the bump with the limits given in Fig.~1 in
Ioka et al. (2005).

\smallskip

As an alternative, the bump around $T_0+1.1\times 10^5$\,s may be
interpreted as due to late internal shocks.  Although this is unusual, other
GRBs have exhibited some late X-ray flares (up to $10^5$\,s) which were
interpreted as due to internal shocks (e.g. GRB 050202B, Falcone et al. 2006
and GRB 050724, Campana et al. 2006c). The quality of the PC data around the
X-ray bump do not allow us to rule out this interpretation.

\section{Conclusion}

GRB 050822 is an XRF showing a complex X-ray light curve: i) an initial steep
decay with three major X-ray flares; ii) a flat decay from $T_0+800$\,s to
$T_0+1.7\times 10^4$; iii) a long and steeper decay up to $T_0+3\times
10^6$\,s with a X-ray bump around $T_0+1.1\times 10^5$\,s.  

We argue that the three X-ray flares observed during the initial steep decay
are likely to be produced by internal processes, and that the global decay is
likely to be the tail of the prompt emission. We showed that the energy peak
of the spectrum for the flare peaking around $T_0+420$\,s is in the XRT energy
band and shifts to lower energy with time. For the flares peaking at
$T_0+131$\,s and $T_0+236$\,s, we showed that $E_p$ is likely to be close to
or less than the lower end of the XRT energy band.

Interestingly, the flare F$_3$ is alternatively well fit by a black-body $+$
power-law (BB-PL) model. We then proposed that the flare F$_3$ may be
produced by photospheric emission (involving Comptonisation) for a shell of
matter moving at a mildly relativistic speed.

We stress that the spectral hardening seen around $\sim T_0+700$\,s (close to
the beginning of the flat decay) can be interpreted as a clear indication of
the emergence of the forward-shock emission. We showed that the emission after
$T_0+700$\,s may then be produced in a site different from that producing
the prompt emission, since the deceleration radius should be larger than
$4\times 10^{16}$\,cm in the case of a CBM afterglow model.

The flat-to-steep decay can then be interpreted as being the afterglow, the
flat part corresponding to a phase of energy injection.  The null detection of
a jet break up to $T_0+3\times 10^6$\,s in the X-ray light curve can be
understood: i) if the jet seen on-axis is uniform with a large opening angle
($\theta > 20^\circ$); ii) if the jet is a structured Gaussian-like jet with
the line-of-sight outside the bright Gaussian core. We note that the same
models were also invoked in the case of GRB 050416A, which is an XRF (Mangano
et al. 2006) to explain the null detection of a jet break in the light-curve.
In both scenarios, the late X-ray bump around $T_0+1.1\times 10^5$\,s could be
produced by internal shocks, implying very late activity of the central
source or it could be produced by inhomogeneity in the blast-wave or by energy
injection.

\bigskip


OG, KPA, MRG, APB, JPO gratefully acknowledge PPARC funding.
DNB, AF, JR and DCM are funded through NASA contract NAS5-00136.


\end{document}